\documentclass[12pt,nofootinbib,aps,amsmath,amssymb,preprint,showpacs]{revtex4}

\usepackage[T1]{fontenc} \usepackage[latin1]{inputenc}
\usepackage{amsmath,color,ulem} \usepackage{graphicx}
\usepackage{epsf,bm} \usepackage{amssymb}

\usepackage{tabulary}
\newcolumntype{K}[1]{>{\centering\arraybackslash}p{#1}}
\newcommand\underrela[2]{\mathrel{\mathop{#2}\limits_{#1}}}
\usepackage{multirow}
\begin{document}

\title{Super slowing down in the bond-diluted Ising model}
\author{Wei Zhong$^\dagger$} 
\email{w.zhong1@uu.nl}
\author{Gerard T. Barkema$^\dagger$}
\author{Debabrata Panja$^\dagger$}
 
\affiliation{
 $^\dagger$Department of Information and Computing Sciences, Utrecht University, Princetonplein 5, 3584 CC Utrecht, The Netherlands\\
}

\date{\today} 

\begin{abstract} 
In models in statistical physics, the dynamics often slows down tremendously near the critical point.
Usually, the correlation time $\tau$ at the critical point increases with system size $L$ in power-law 
fashion: $\tau \sim L^z$, which defines the critical dynamical exponent $z$. We show that this also holds
for the 2D bond-diluted Ising model in the regime $p>p_c$, where $p$ is the parameter denoting the bond concentration, 
but with a dynamical critical exponent $z(p)$ which shows a strong $p$-dependence.
Moreover, we show numerically that $z(p)$, as obtained from the autocorrelation of the total magnetisation, 
diverges when the percolation threshold $p_c=1/2$ is approached:
$z(p)-z(1) \sim (p-p_c)^{-2}$. We refer to this observed extremely fast increase of the correlation time with size 
as {\it super slowing down}. Independent measurement data from
the mean-square deviation of the total magnetisation, which
exhibits anomalous diffusion at the critical point, supports this result.
\end{abstract}

\maketitle
 
\section{Introduction \label{sce1}}

The Ising model has proven to be a staple model in physics for
studying phase transitions and critical phenomena
\cite{stanley,hohen}. The model was originally conceived to provide a
theoretical understanding of the existence of a Curie temperature for
``pure'' ferromagnetic materials; purity here refers to the fact that
all throughout the material, every lattice site contains a spin, and
every spin interacts uniformly with the surrounding ones. From that
point of view, it can be argued that in nature pure materials are
rare, i.e., impurities are by and large inevitable.

In the Ising model, the impurities have been implemented in terms of
randomly placed nonmagnetic spins (site-diluted Ising model)
\cite{balle1,balle2,martins,kim,kenna} or missing interactions (bond-diluted
Ising model) \cite{shalaev,jayap,loh,hasen,rivera}. The inclusion of
any kind of randomness into the system can have significant effects on
its critical properties \cite{alba}. For instance, a new universality
class was found in the three-dimensional bond-diluted Ising model
\cite{hasen2,fytas}, and complex logarithmic corrections for the
equilibrium properties were observed \cite{kenna,wang}. Moreover, a
large number of novel crossover behaviour between pure and percolating
Ising systems have been found
\cite{heuer,mouri,tomita,jassen2,heuer2,domany}. Also, the dynamics at the percolation threshold is discussed in Refs. \cite{harris0,jain} and the dynamical exponent for spin systems with random dilution, or randomness in the coupling constants has been considered in Refs. \cite{prud,paris,ivaney,henk,lipp}. Despite these
advances, dynamical properties of the bond-diluted Ising model (i.e.,
as a function of bond concentration $p$) remains poorly studied.

In this paper, we take on studying the slowing down of the dynamics of
the total magnetisation autocorrelation function at the critical
temperature $T_c(p)$ in the square $(L\times L)$ two-dimensional
bond-diluted Ising model with Monte Carlo simulations. To this end,
using the Binder cumulant, we first measure $T_c(p)$ at several values
of $p$. We then turn to the calculation of $z(p)$ for several $p>p_c$
from the total magnetisation autocorrelation function: by collapsing
this autocorrelation function to a reference curve, we calculate the
relative terminal exponential decay time $\tau[T_c(p)]$ for the
correlation function. Thereafter, by fitting this data as
$\tau[T_c(p)]\sim L^{z(p)}$, we directly extract $z(p)$. As
$p\rightarrow p_c^+$, we empirically find that the dynamical exponent
$z(p)$ increases continuously as $z(p)-z(1)\sim(p-p_c)^{-2}$, with
$z(1)=2.1665(12)$ the dynamical critical exponent of the ordinary
Ising model \cite{night}.

Further, we also consider the mean-square deviation (MSD) of the total
magnetisation $M$ of the model, which for $p=1$ has been shown to
exhibit anomalous diffusion as $\langle\Delta M^2(t)\rangle\sim
t^\alpha$ with the anomalous exponent
$\alpha=\displaystyle{\frac{\gamma}{\nu z(1)}}$ \cite{zhong1}, with
$\gamma=7/4$ and $\nu=1$. Given that the equilibrium critical
exponents $\gamma$ and $\nu$ are numerically nearly independent of $p$
for $p\geq 0.6$ \cite{hadji}, combined with values for $z(p)$ as
obtained through the terminal relaxation time for different $p$, the
various MSD-curves of the total magnetisation are collapsed on top of
each other with a $p$-dependent shift factor $\mathcal{G}(p)$ via
$\log\left( \langle \Delta M^2\rangle /
L^{2+\gamma/\nu}\right)/\alpha(p) \sim \log\left(
t/L^{z(p)}\right)+\log\left( \mathcal{G}(p)\right)/\alpha(p)$, with
$\alpha(p)=\gamma(p)/[\nu(p)\, z(p)]$. The result reveals that the
magnetisation indeed experiences anomalous diffusion at the critical
point, for a range of dilution $p>p_c$.  The collapse of the MSD of
the magnetisation confirms that the measured values of $z(p)$ are
correct.

The paper is organised as follows. In Sec. \ref{sec2} we introduce the
2D bond-diluted Ising model and measure its critical temperature at
several values of $p$. In Sec. \ref{sec3} we obtain the dynamical
exponent $z(p)$ from the total magnetisation autocorrelation
function. In Sec. \ref{sec4}, we confirm $z(p)$ values from the
exponent of anomalous diffusion of the MSD of the total
magnetisation. We conclude the paper in Sec.  \ref{sec5}.

\section{Bond-diluted Ising model and its critical temperature\label{sec2}}

We consider the two-dimensional (2D) bond-diluted Ising model on an
$L\times L$ square lattice with periodic boundary conditions. For this
model the Hamiltonian, without an external field, is given by
\begin{equation}
{\cal H}=-\sum_{\langle ij \rangle} J_{ij}s_i s_j,
\end{equation}
where $s_i=\pm 1$ is the spin residing at site $i$, $\langle ij \rangle$
denotes the sum running over all nearest neighbour sites, and the coupling
constant $J_{ij}$ is given by the distribution function
\begin{equation}
P(J_{ij})=p \delta(J_{ij}-1)+(1-p)\delta(J_{ij}),
\label{eqJ}
\end{equation}
with $p$ being the bond concentration ($0\le p\le1$). The function
(\ref{eqJ}) simply means that the value of $J_{ij}$ is $1$ with probability
$p$, and $0$ otherwise.

\begin{table}[h] 
\begin{tabular}{K{4mm}|K{1.6cm}|K{1.6cm}|K{1.6cm}|K{1.6cm}|K{1.6cm}|K{1.6cm}|K{1.6cm}|K{1.6cm}|K{1.6cm}} 
$p$ & $0.9$  & $0.85$ & $0.8$  & $0.75$  & $0.7$ & $0.65$  &  $0.6$  & $0.58$ & $0.55$\\
\hline\hline
  $N$  & $2000$ & $5000$ & $20000$ & $20000$ & $20000$ & $20000$ & $200000$ & $200000$ & $400000$\\
\hline
  $T_c$  & $1.956(10)$ & $1.804(10)$  & $1.650(20)$ & $1.472(20)$ & $1.310(25)$ & $1.141(30)$ & $0.951(20)$ & $0.869(25)$ & $0.727(40)$\\
  \hline
\end{tabular} 
\caption{Number of samples $N(p)$ used to measure $T_c(p)$, and the
  simulation results for $T_c(p)$ (including error bars) for different bond
  concentrations $p$.  \label{tab1}} 
\end{table}

\begin{figure*}[htb]
\includegraphics[width=0.45\linewidth]{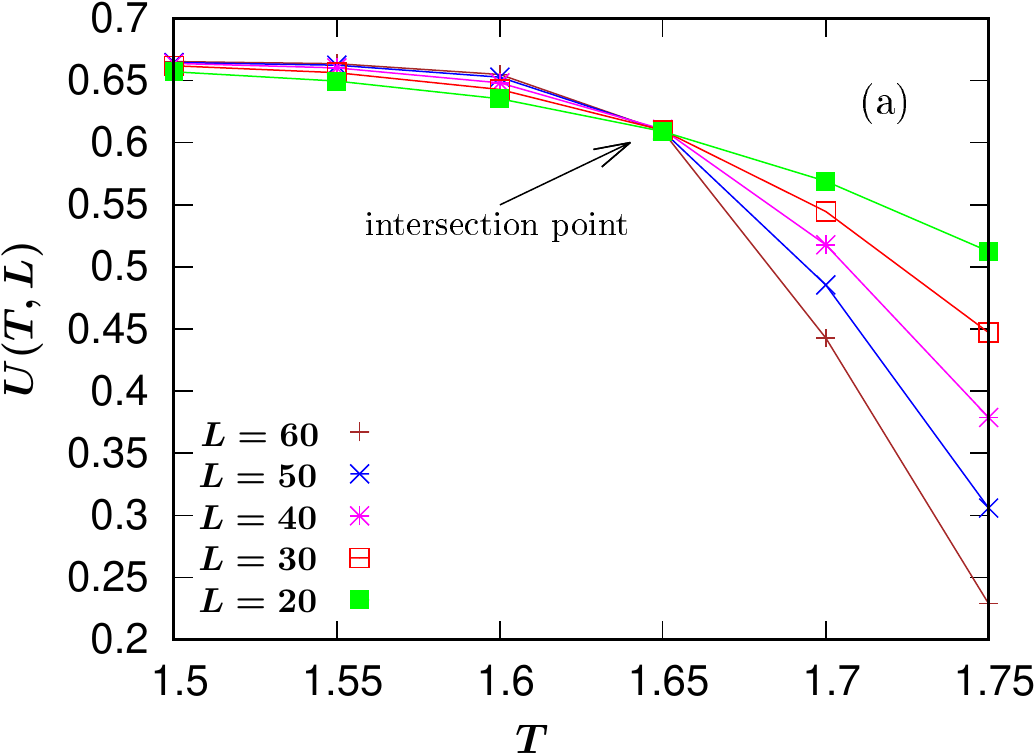}
\hspace{3mm}
\includegraphics[width=0.45\linewidth]{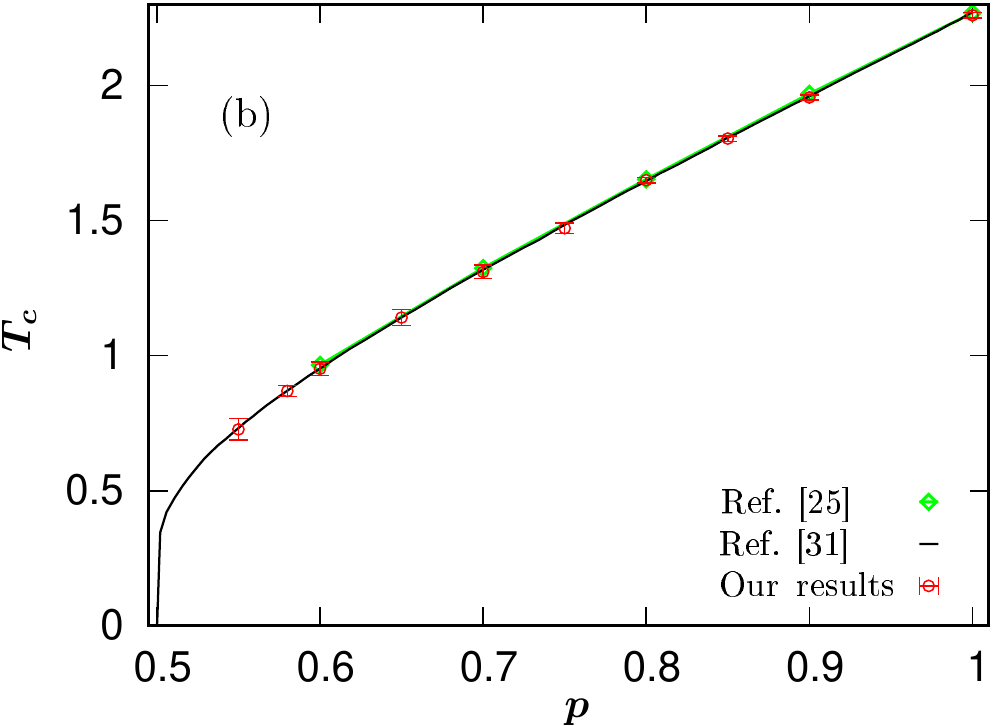}
\caption{(a) Example calculation of $T_c(p)$ for the
  2D bond-diluted Ising model for $p=0.8$ using Binder cumulant. The
  $x$-value of the intersection point indicates that
  $T_c(p=0.8)=1.650\pm 0.020$. (b) Critical temperatures for different 
  values of $p$, as noted in Table \ref{tab1}. Our results match those
  of Refs. \cite{ohzeki,hadji} very well.}
  \label{binder_cumu}
\end{figure*}

For the pure Ising model ($p=1$), there is
a second-order phase transition at $T_c(1)=2/\ln(1+\sqrt{2})$
\cite{binney}. When $p$ reaches the percolation threshold
$p_c=1/2$, its critical temperature decreases to zero: $T_c(p_c)=0$
\cite{nishi}. To determine $T_c(p)$ for in-between
values of $p$, we use the Binder cumulant. It is defined as
\cite{binder}
\begin{equation}
U(T,L)=1-\frac{\langle M^4\rangle}{3 \langle M^2\rangle^2},
\end{equation}
where $\langle M^4\rangle$ and $\langle M^2\rangle$ are the thermal
averages of the fourth and second moments of the total magnetisation
$M=\displaystyle{\sum_{i=1}^{L\times L}} s_i$. For each value of $p$,
the curves of $U(T,L)$ plotted vs. $T$ for various values of $L$
intersect at a fixed point, which determines the critical
temperature. The process is illustrated in Fig.  \ref{binder_cumu}(a).

We perform Monte Carlo simulations using the Wolff algorithm
\cite{wolff,newman} to calculate $T_c(p)$. Running many independent
samples provide us with fairly accurate values of these critical
temperatures, as noted in Table \ref{tab1}. In Fig.
\ref{binder_cumu}(b) we show that the values for $T_c(p)$ obtained
this way match very well with those in Refs. \cite{ohzeki,hadji}.

\section{Dynamical exponent for different values of $p$ \label{sec3}}

Having obtained the critical temperatures for a number of $p$-values
as per above, in this section, we measure the total magnetisation
autocorrelation function $\langle M(t)\cdot M(0)\rangle$. In order to
do so, we first run $2\times 10^6$ Wolff Moves to thermalise the
system. Subsequently, we evolve the system following Glauber dynamics,
i.e. spin flips are proposed at random locations, and accepted with
the Metropolis acceptance probability.  Time is measured in terms of
attempted Monte Carlo moves, since every spin attempts to flip
statistically once per unit time. As we continue to do so, we keep
taking snapshots of the full system at regular intervals over a total
time of $2\times 10^7$ attempted Monte Carlo moves per lattice site,
and correspondingly compute the total magnetisation $M$ at every
snapshot. This leads us to $\langle M(t) \cdot M(0)\rangle$. For
different values of $p$, we run $500$ to $2000$ independent
simulations to achieve decent accuracy. We vary the system size from
$10$ to $40$.

\begin{figure*}[htb]
  \includegraphics[width=0.5\linewidth]{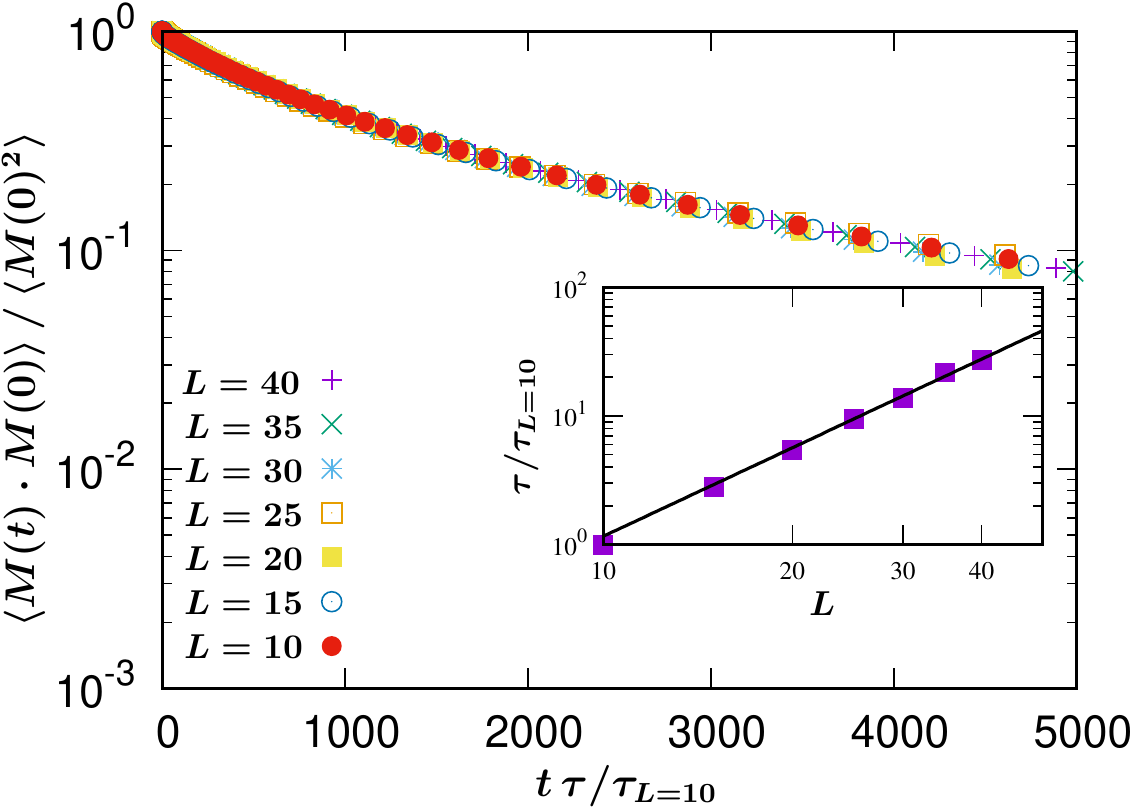}
  \caption{The collapse of $\langle M(t)\cdot M(0)\rangle/\langle
    M(0)^2\rangle$ as a function of $t\tau[T_c(p)]/\tau[T_c(p)]_{L=10}$
    for $p=0.8$. The system size varies from $10$ to $40$. Inset:
    correspondingly, $\tau[T_c(p)]/\tau[T_c(p)]_{L=10}$ as a function 
    of $L$. The dynamical exponent is obtained by fitting these data 
    as $\tau[T_c(p)]/\tau[T_c(p)]_{L=10}\sim L^{z(p)}$. The solid
    line corresponds the function $y=x^{2.285}$. From this we obtain
    $z(0.8)\approx 2.285$.}
  \label{corr_term}
\end{figure*}

For a given value of $p$ and the corresponding critical temperature
$T_c(p)$, we collapse all the curves for the normalised total
magnetisation autocorrelation function $\langle M(t) \cdot
M(0)\rangle/\langle M(0)^2\rangle$ to a reference curve ($L=10$). This
allows us to compute the ratio of the terminal decay times
$\tau[T_c(p)]/\tau[T_c(p)]_{L=10}$. Fig. \ref{corr_term} demonstrates
this procedure for $p=0.8$: with a properly chosen value of
$\tau[T_c(p)]/\tau_{L=10}[T_c(p)]$, the $\langle M(t) \cdot
M(0)\rangle/\langle M(0)^2\rangle$ data for different system sizes
collapse on the curve corresponding to $L=10$.
 
\begin{figure*}[htb]
\includegraphics[width=0.5\linewidth]{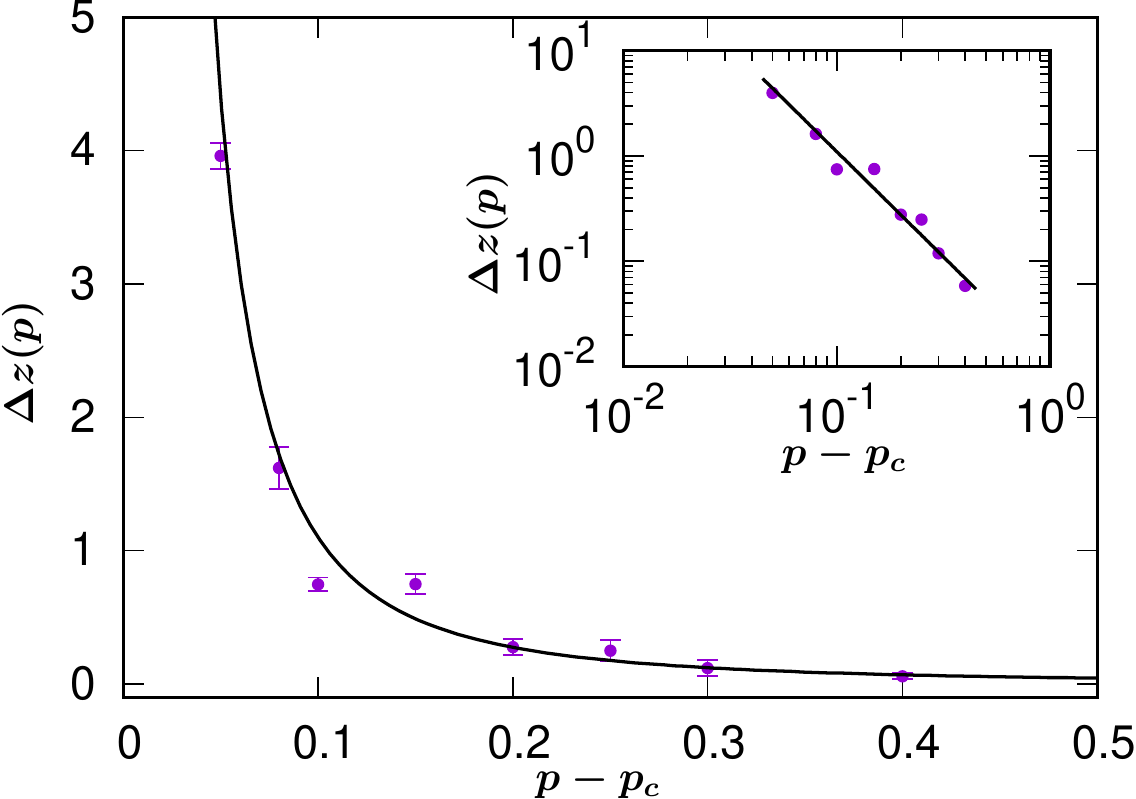}
\caption{The dynamical exponent difference
  $\Delta z(p)=z(p)-z(1)$ as a function of $p-p_c$, where
  $z(1)=2.1665(12)$ \cite{night} is the dynamical exponent for the
  pure $2D$ Ising model. The result implies that
  $z(p)\rightarrow\infty$ as $p\rightarrow p_c^+$.}
  \label{dyna_expo}
\end{figure*}

Further, given our argument in Appendix A that $L$ is the
characteristic length scale for $L\ge10$ for the 2D bond-diluted Ising
model when $p\ge0.6$, we have, at the critical temperature,
\begin{equation}
\tau(p)/\tau_{L=10}(p) \sim L^{z(p)}.
\label{tau}
\end{equation}
By plotting the $\tau(p)/\tau_{L=10}(p)$ data (inset Fig.
\ref{corr_term}), we extract $z(p)$. The results from this exercise
for several values of $p$ are shown in Fig. \ref{dyna_expo}.
Numerically, therein we find that 
\begin{equation}
\Delta z(p)=z(p)-z(1)\sim (p-p_c)^{-2}\quad \mbox{for} \quad
p_c < p <1,
\label{ezscaling}
\end{equation}
where $z(1)=2.1665(12)$ \cite{night} is the dynamical exponent for
the pure 2D Ising model.

Based on concepts of renormalisation, we anticipated that away from
the percolation threshold $p_c$, the correlation time as a function of
system size would show a crossover from the behaviour for the
bond-diluted Ising model at small system sizes to that of the ordinary
Ising model at large system sizes, with a crossover size that diverges
if $p_c$ is approached.  Instead, we find a fairly clean power-law
behaviour of the correlation time for all system sizes, with a single
exponent $z(p)$ that varies strongly with $p$. As a function of bond
dilution $p$, the dynamical exponent $z(p)$ increases monotonically
when $p$ decreases from $p=1$ to $p=p_c$. Moreover, the numerical
results suggest that $z(p)$ will become infinitely large when $p$
approaches the percolation threshold $p\rightarrow p_c^+$, i.e., the
dynamics of the system gets extremely slow as $p\rightarrow p_c^+$, a
phenomenon we term as ``super slowing down''.

\section{Anomalous diffusion of the total magnetisation \label{sec4}}

To confirm the observed behaviour of super slowing down [i.e., the Eq.
(\ref{ezscaling})] for the bond-diluted Ising model by means of
independent measurements, we now focus on the mean-square deviation
(MSD) of the magnetisation as a function of time $t$ as 
\begin{equation} 
\langle \Delta M^2\rangle=\langle [M(t)-M(0)]^2\rangle. 
\end{equation}

At short times ($t \approx 1$), changes in $M$, occurring due to
random thermal fluctuations of individual spins, are uncorrelated;
hence $\langle \Delta M^2\rangle\sim L^2 t$ for 2D Ising model. At
long times, $t\gtrsim L^{z(p)}$, we expect
$\langle M(t)\cdot M(0)\rangle=0$, meaning that \begin{equation}
  \langle \Delta M^2\rangle\underrela{t\gtrsim L^{z(p)}}{=}2\langle
  M(t)^2\rangle/\sim L^{2+\gamma(p)/\nu(p)}. 
\end{equation}
If we assume that the MSD is given by a simple power law
in the intermediate time regime ($1 \gtrsim t \gtrsim L^{z(p)}$ ),
then we obtain
\begin{equation}
\langle \Delta M^2\rangle\sim L^{2+\gamma(p)/\nu(p)} \left(t/L^{z(p)}\right)^{\alpha(p)},
\label{msd0}
\end{equation}
where $\alpha(p)=\displaystyle{\frac{\gamma(p)}{\nu(p) z(p)}}$.
For the pure Ising model in two dimensions ($p=1$), we have shown
that \cite{zhong1}
\begin{equation}
\langle \Delta M^2\rangle/ L^{2+\gamma/\nu} = f(t/L^{z}),
\label{msd}
\end{equation}
where $\gamma=7/4$ and $\nu=1$ are two equilibrium critical exponents.
Here $f(x)$ is a scaling function such that
$\displaystyle{\lim_{x\rightarrow0}}f(x)\sim x^{\gamma/(\nu z)}\approx
x^{0.81}$, and $f(x)$ saturates for $x\gtrsim1$. Indeed, given that
$\gamma(p)$ and $\nu(p)$ are nearly independent of $p$ when $p \geq
0.6$ \cite{hadji} (see also in Appendix B), if the scaling relation
(\ref{msd}) also continues to hold for values of $p$ other than unity,
then we can use it to obtain independent confirmation for the super
slowing down (\ref{ezscaling}). We demonstrate this below by focusing
on $p\geq 0.6$.
\begin{figure}[h]
\includegraphics[width=0.55\linewidth]{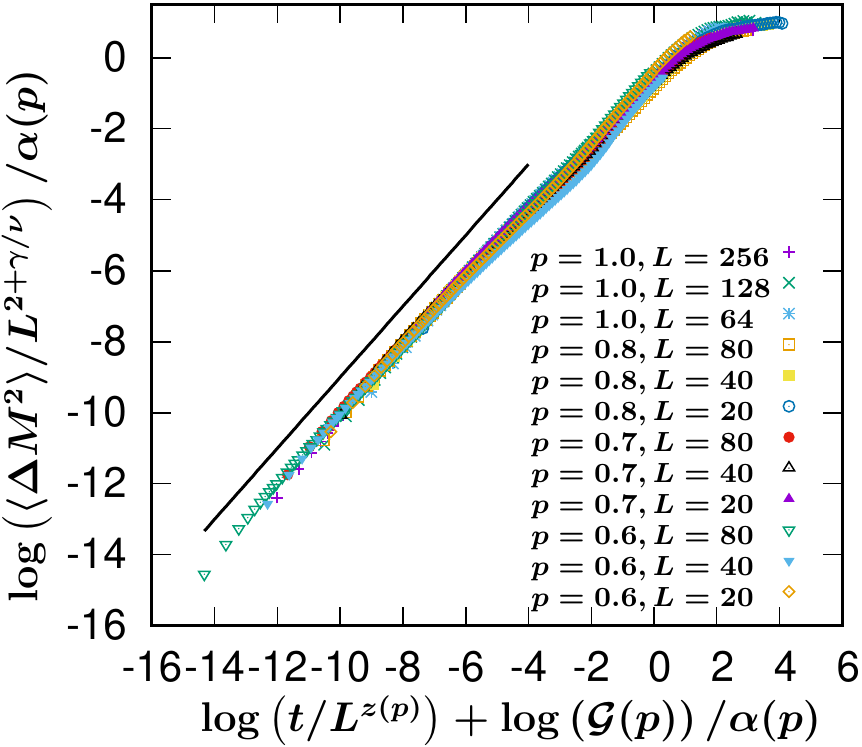}
 \caption{The collapse of the mean-square displacement of the total
   magnetisation via $\log(\langle \Delta M^2\rangle /
   L^{2+\gamma/\nu})/\alpha(p) \sim \log\left(
   t/L^{z(p)}\right)+\log\left( \mathcal{G}(p)\right)/\alpha(p)$,
   where the obtained values of $z(p)$ from the last section are
   employed and $\mathcal{G}(p)$ is a $p$-dependent shift factor. The slope of the solid line is unity. It
   confirms that the MSD of the total magnetisation experiences
   anomalous diffusion at $T_c(p)$ and the values of $z(p)$ is
   increasing when $p\rightarrow p_c+$.}
  \label{fig_msd}
\end{figure}

Since in the previous section, we have obtained the values of $z(p)$
for different $p$, here, we describe the MSD of the total
magnetisation by modifying Eq. (\ref{msd0}) as
\begin{equation}
\langle \Delta M^2\rangle / L^{2+\gamma/\nu}\sim \mathcal{G}(p)\left(t/L^{z(p)}\right)^{\alpha(p)},
\label{msd1}
\end{equation}
where $\mathcal{G}(p)$ is a $p$-dependent shift factor. We
take $logarithm$ of both sides of Eq. (\ref{msd1}) to write
\begin{equation}
\log\left( \langle \Delta M^2\rangle / L^{2+\gamma/\nu}\right)/\alpha(p) \sim \log\left( t/L^{z(p)}\right)+\log\left( \mathcal{G}(p)\right)/\alpha(p).
\label{msd2}
\end{equation}

Suppose we choose the MSD of the total magnetisation for the normal
Ising model as the reference [means that we set $\mathcal{G}(1)\equiv
  1$], if the values of $z(p)$ obtained from the last section are
correct, then with these $z(p)$ values and the shift factor ${\mathcal
  G}(p)$, the MSD of the total magnetisation for different $p$ can be
made to collapse onto the data for $p=1$ via Eq. (\ref{msd2}).

In order to obtain the $\langle\Delta M^2(t)\rangle$ data, once again,
we first thermalise the system with $2 \times 10^6$ Wolff moves, then
measure $\langle \Delta M^2\rangle$ in a further simulation over $2
\times 10^7$ attempted Monte Carlo moves per lattice site. We use
three different system sizes: $L = 20, 40, 60$ for every value of $p$.
 
Figure \ref{fig_msd} implies that by using the values of $z(p)$
obtained from the last section, indeed for different $p$, the MSD of
the magnetisation can be collapsed onto the data for $p=1$ via
Eq. (\ref{msd2}). It confirms that the MSD of the total magnetisation
experience anomalous diffusion at $T_c(p)$ and $z(p)$ values obtained
from the terminal relaxation time are correct.
  
In summary, with two different methods, we have shown that $z(p)$ is
diverging when $p\rightarrow p_c^+$, i.e., the dynamics of the system
is getting extremely slow when we reduce the bond concentration to its
percolation threshold.  We do not have a quantitative explanation for
this behaviour. That said, it might arise from the fact that the
fraction of `unhappy' bonds (active bonds between sites with opposing
spin values) at the critical temperature decreases to zero if $p_c$ is
approached, thereby removing the energetic contribution of restoring
forces; we provide some measurements for this in Appendix C.

\section{Discussion}

In this paper, we study the critical dynamical exponent $z(p)$ for
the 2D bond-diluted Ising model with bond concentration $p$. We first
measure the critical temperature $T_c(p)$ for different bond
concentrations $p$ using the Binder cumulant. We then calculate the
relative values of the terminal decay time $\tau$ by collapsing the
total magnetisation autocorrelation function to a reference value,
from which we obtain $z(p)$ using the relation
$\tau\sim L^{z(p)}$.

We find that $z(p)$ increases when $p\rightarrow p_c^+$ as a power-law
$z(p)-z(1)\sim(p-p_c)^{-2}$, which we refer to as super slowing
down. We confirm this result from independent measurements of the
MSD of the total magnetisation that exhibits anomalous diffusion.

Our results indicate that $z(p)\rightarrow\infty$ as $p\rightarrow
p_c^+$. This leaves us with the interesting question: what happens to
$z(p)$ when $p<p_c$? We plan to explore this in future.

\section*{Acknowlegement \label{sec5}}

W.Z. acknowledges financial support from the China Scholarship Council (CSC).

\section*{Appendix A: Relevant length scale for critical phenomena of the
  bond-diluted Ising model\label{apb}}

\begin{figure*}[htb]
  \includegraphics[width=0.4 \linewidth]{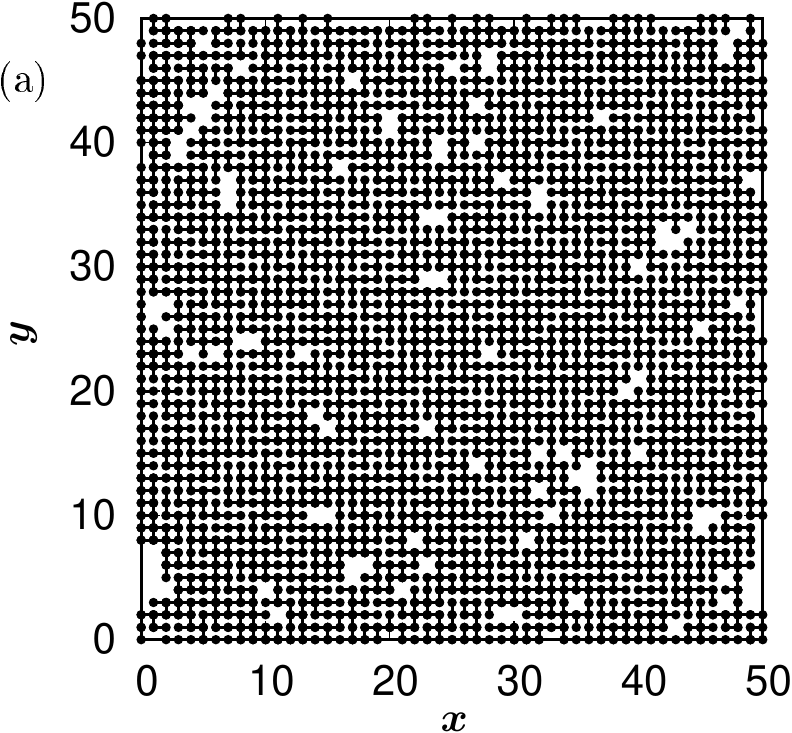}
  \hspace{3mm}
  \includegraphics[width=0.54\linewidth]{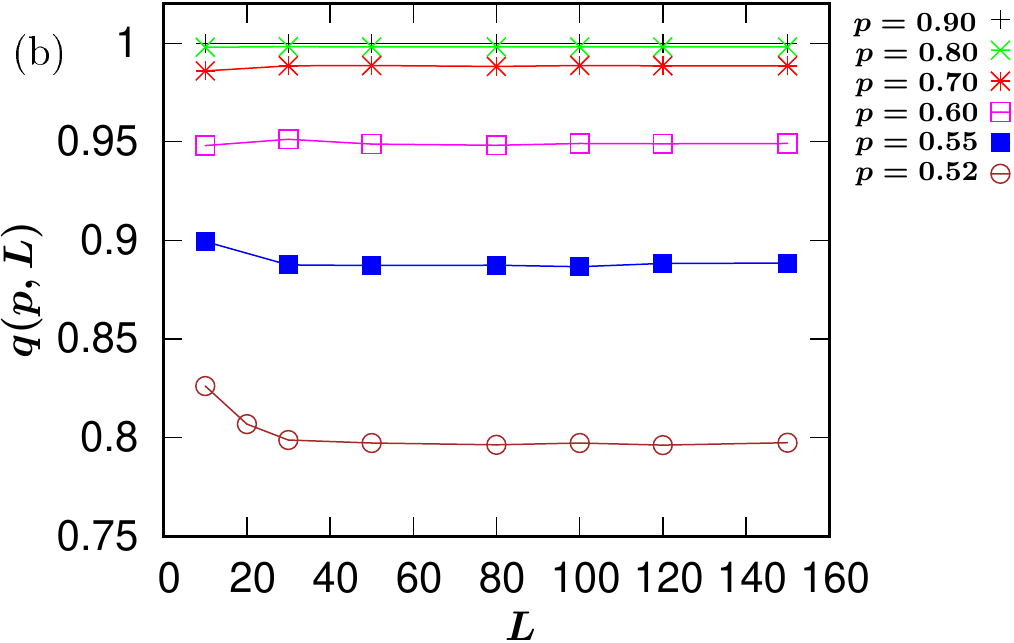}
  \caption{(a) A snapshot of the biggest knot cluster for $L=50$ and
    $p=0.6$. Spins that are not belong to this cluster are represented
    by cavities. (b) Plot of $q(p,L)=\langle S(p,L)
      \rangle/L^2$ for various values of $p$ and $L$: for $p\ge0.6$,
      $q(p,L)$ is independent of $L$ for $L\ge10$. }
  \label{fig_q}
\end{figure*}

In the pure 2D Ising model of dimension $L\times L$, the only relevant
length scale for critical phenomena is $L$. For the bond-diluted Ising
model there are other length scales, for instance corresponding to the
size of the biggest cluster, $S(p,L)$. Here, a cluster is defined is
by the set of spins such that there is at least one continuous
(bond-following) path from every spin in the cluster to every other
spin in the same cluster. We define the size of the cluster by the
total number of spins belonging to the cluster [thus $S(p,L)$ is the
  number of spins in the biggest cluster for an $L\times L$ system
  with bond concentration $p$].

In Fig. \ref{fig_q}(b) the quantity considered is $q(p,L)=\langle
S(p,L)\rangle/L^2$. If this quantity is independent of $L$ then it
means that there is no difference between the two differently defined
length scales (apart from a scaling factor).  For each result, we have
generated $500$ samples. We see in Fig.  \ref{fig_q}(b) that for $p\ge
0.6$, $q(p,L)$ is independent of $L$ for $L\ge 10$. This means that
for the range of dilution $p\ge0.6$ used in this paper, we can use $L$
as the relevant length scale for critical phenomena provided $L\ge
10$.
  
\section*{Appendix B: Equilibrium critical exponents $\nu$ and $\gamma$ \label{apa}}

In this Appendix, we show that the equilibrium critical exponents
$\nu$ and $\gamma$ for the bond-diluted Ising system with $p\geq 0.6$
are numerically indistinguishable from their values in the pure Ising
model.

We note here that  according to the Harris criterion \cite{harris}, if the correlation length critical exponent $\nu$ fulfills the inequality $\nu\geq 2/d$ where $d$ is the spatial dimensionality, then disorder does not affect the critical behavior. For the $2D$ Ising model, $\nu = 1$ is marginal, which translates into logarithmic corrections to some critical exponents. In the pure Ising model, the exponents  $\gamma$ and $\nu$ do not show logarithmic corrections, and our numerical results shown in this Appendix indicate that the ratio of $\gamma$ and $\nu$ is unchanged in the regime we studied, for $p \geq 0.6$, without logarithmic corrections. Also, the Binder cumulant does not show logarithmic corrections. This is not obvious, and in fact there are reports of logarithmic corrections to the equilibrium properties of the diluted spin systems \cite{kenna,aar,mar}.  We cannot rule out the possibility to have logarithmic corrections in the quantities measured by us, as these are difficult to observe in simulations.

Firstly, if we get the values of $T_c$ for different $p$, the Binder
cumulant can be scaled as
\begin{equation}
U(T,L)\sim f(T' L^{1/\nu(p)}),
\end{equation}
which will provide us the value of $\nu(p)$. Here $T'=(T-T_C)/T_C$ is
the reduced temperature.

\begin{figure*}[htb]
\includegraphics[width=0.45\linewidth]{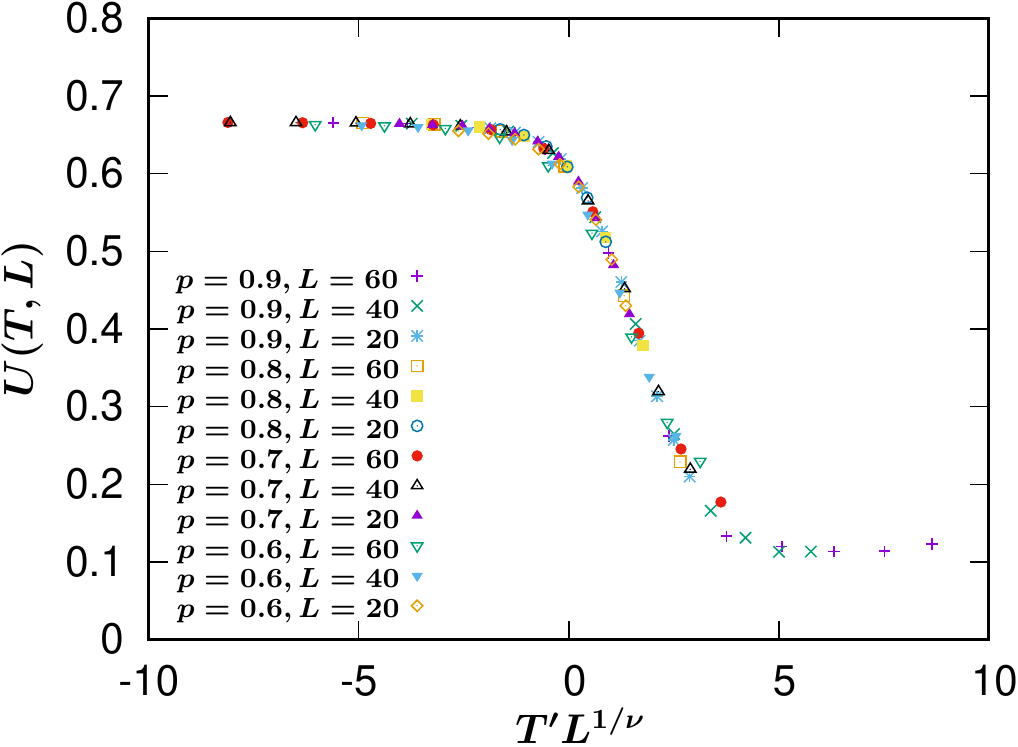}
 \caption{The Binder cumulant for $p\geq 0.6$. The data is collapse as
   $U(T,L)\sim f(T' L^{1/\nu(p)})$ for different $p$ and $L$, where
   $T'=(T-T_C)/T_C$ is the reduced temperature. Here $L=20,40$ and
   $60$ for each bond concentration, and we set all $\nu(p)=1.0$ to
   collapse the data.}
  \label{ap1}
\end{figure*}

In Fig. \ref{ap1}, we collapse the data of $U(L,T)$ for $L=20, 40$ and
$60$ with $\nu(p)\approx1$. It indicates that $\nu(p)$ is numerically
indistinguishable from unity for $p\geq 0.6$.

Next, we turn to measure the magnetic susceptibility $\chi$. For this
simulation, we have used $500$ independent samples for each value of
$p$. It is a well known result \cite{newman} that the susceptibility
can be scaled as
\begin{equation}
\chi L^{-\gamma/\nu} =\tilde{\chi}(T' L^{1/\nu}),
\label{chi}
\end{equation}
where $\tilde{\chi}$ is a dimensionless function.

\begin{figure*}[htbp]
\includegraphics[width=0.42\linewidth]{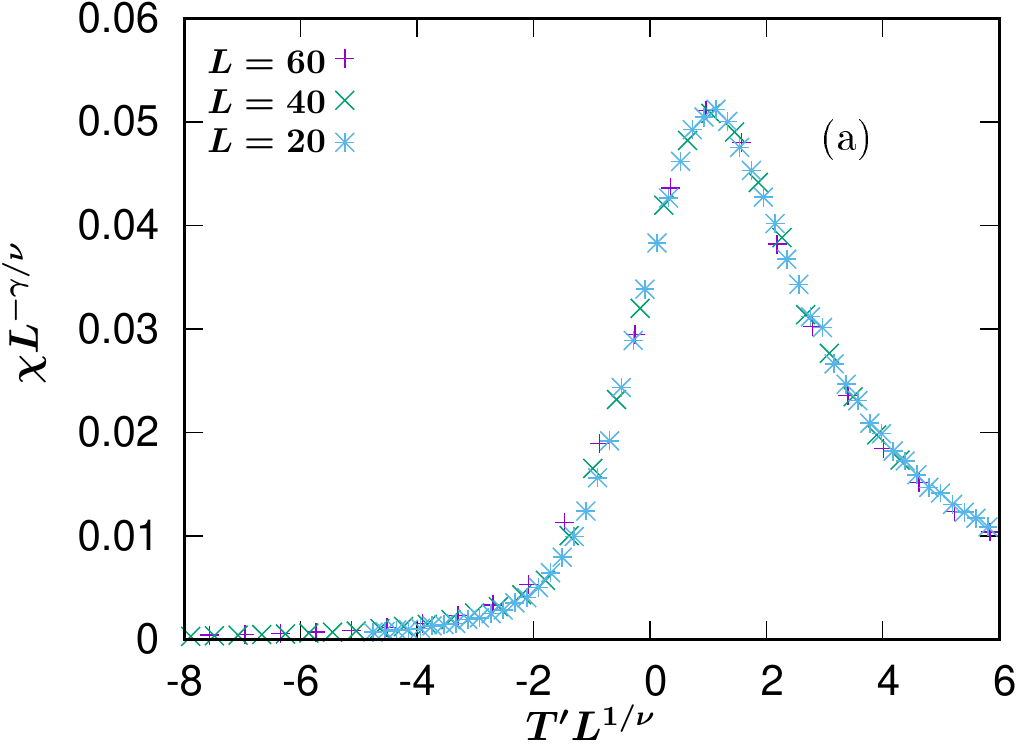}\hspace{3mm}
\includegraphics[width=0.42\linewidth]{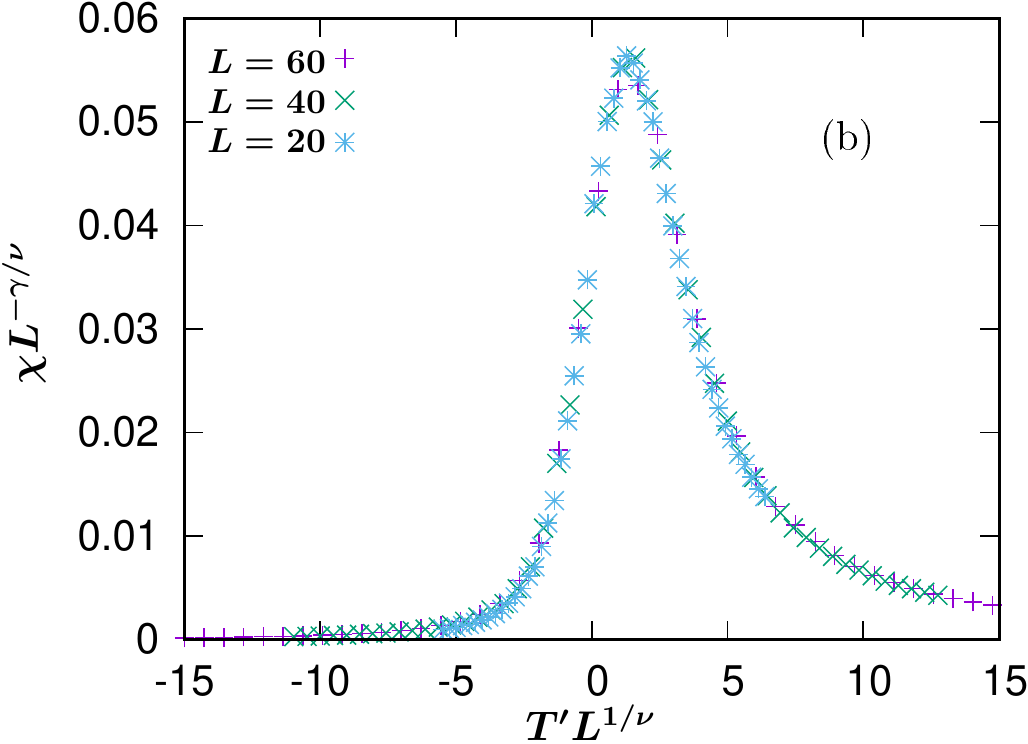}\\
\includegraphics[width=0.42\linewidth]{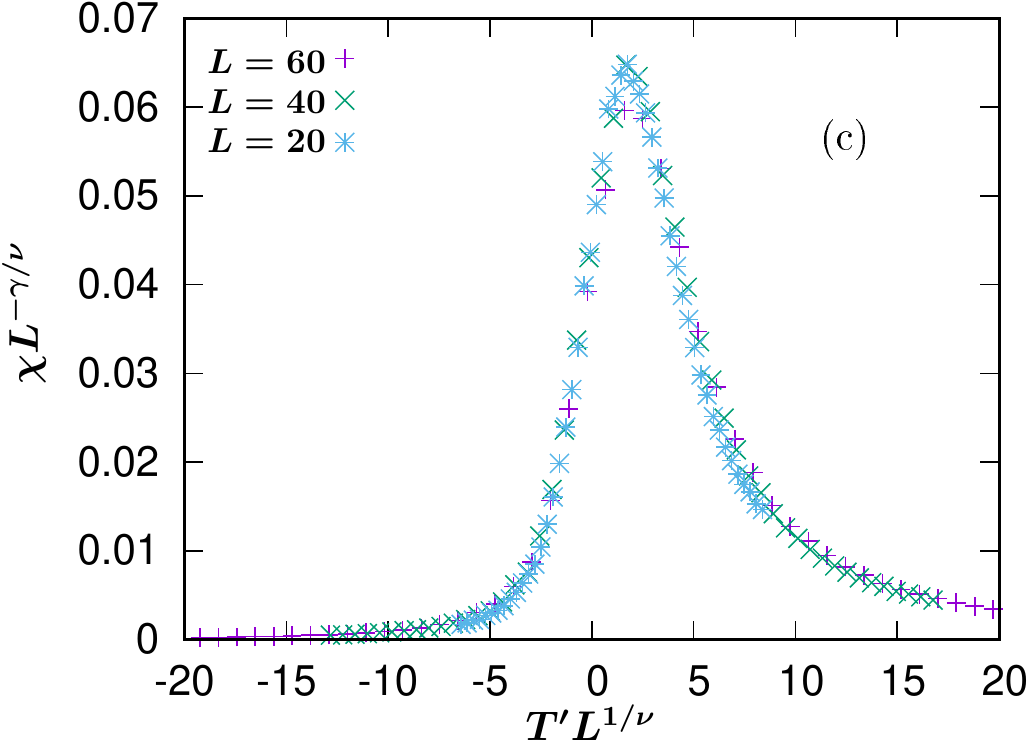}\hspace{3mm}
\includegraphics[width=0.42\linewidth]{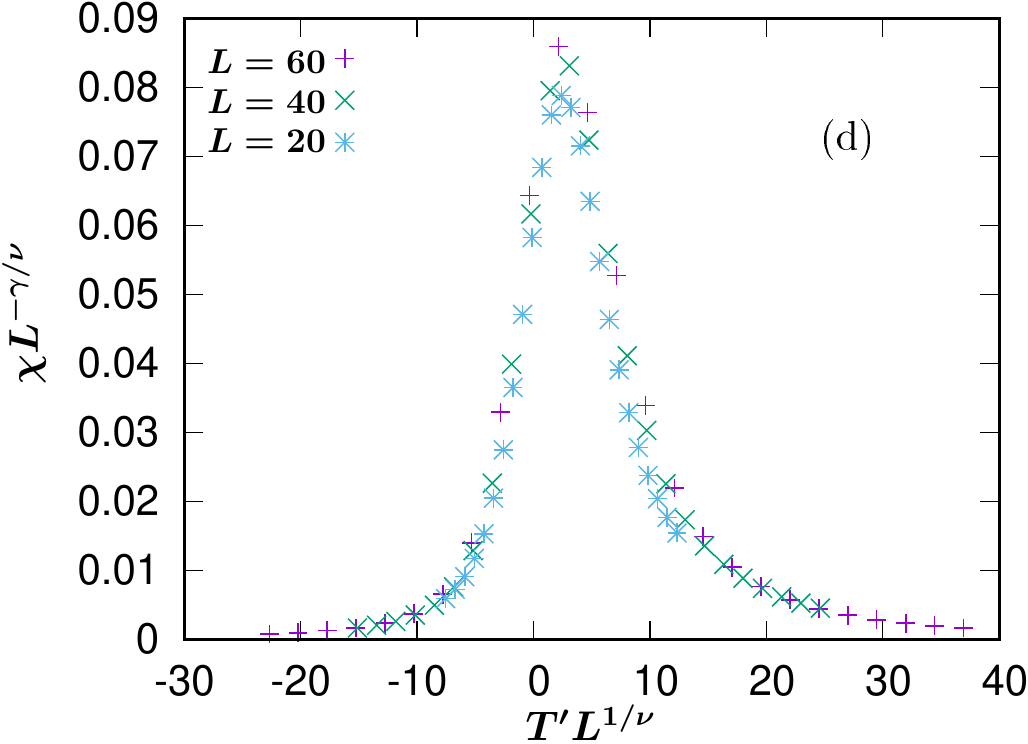}
 \caption{The scaling of magnetic susceptibility as a function of
   reduced temperature: $\chi L^{-\gamma/\nu} =\tilde{\chi}(T'
   L^{1/\nu})$. Here $\tilde{\chi}$ is a dimensionless function, the
   values of $\gamma(p)$ and $\nu(p)$ are chosen to be their values
   for the normal Ising model, i.e., $\gamma(p)=1.75$ and
   $\nu(p)=1$. For figures (a)-(d), the bond concentrations are
   $p=0.9,0.8,0.7$ and $0.6$. The well collapse of all the data
   indicates that both $\gamma$ and $\nu$ are numerically
   indistinguishable for $p\geq 0.6$. }
  \label{ap2}
\end{figure*}

After rescaling the susceptibility using Eq. (\ref{chi}), the data
shown in Fig. \ref{ap2} demonstrate that $\gamma$ is numerically
indistinguishable from $7/4$ for $p\geq 0.6$.

In other words, in this Appendix we have shown that $\gamma$ and $\nu$
are numerically indistinguishable respectively from $7/4$ and unity
for $p\geq 0.6$, confirming the results from Ref. \cite{hadji}.
 
\subsection*{Appendix C: Number of different types of bonds}

In this Appendix, we connect the super slowing down in the $2D$
bond-diluted Ising model with its equilibrium property, i.e., the
ensemble average of the number of `unhappy' bonds, i.e. the number of
interacting nearest-neighbour spins with opposite signs at the
critical temperature.

In the bond-diluted Ising model, we distinguish inactive bonds (with
$J_{ij}=0$), active bonds connecting sites with aligned spins, and
active bonds that connect sites with spins of opposite signs.  For the
active bonds, we denote the numbers of those aligned and nonaligned
spins by $(n_{++}+n_{--})$ and $n_{+-}$, respectively.  Energetically,
$\langle n_{++}\rangle$ and $\langle n_{--}\rangle$ are the bonds that
try to keep the system as it is, and $\langle n_{+-} \rangle$ is
driving spins to flip. If $\langle n_{+-} \rangle$ decreases, then
most of the proposed spin flips will be rejected and the dynamics of
the system will get slower.

\begin{figure*}[htbp]
\includegraphics[width=0.45\linewidth]{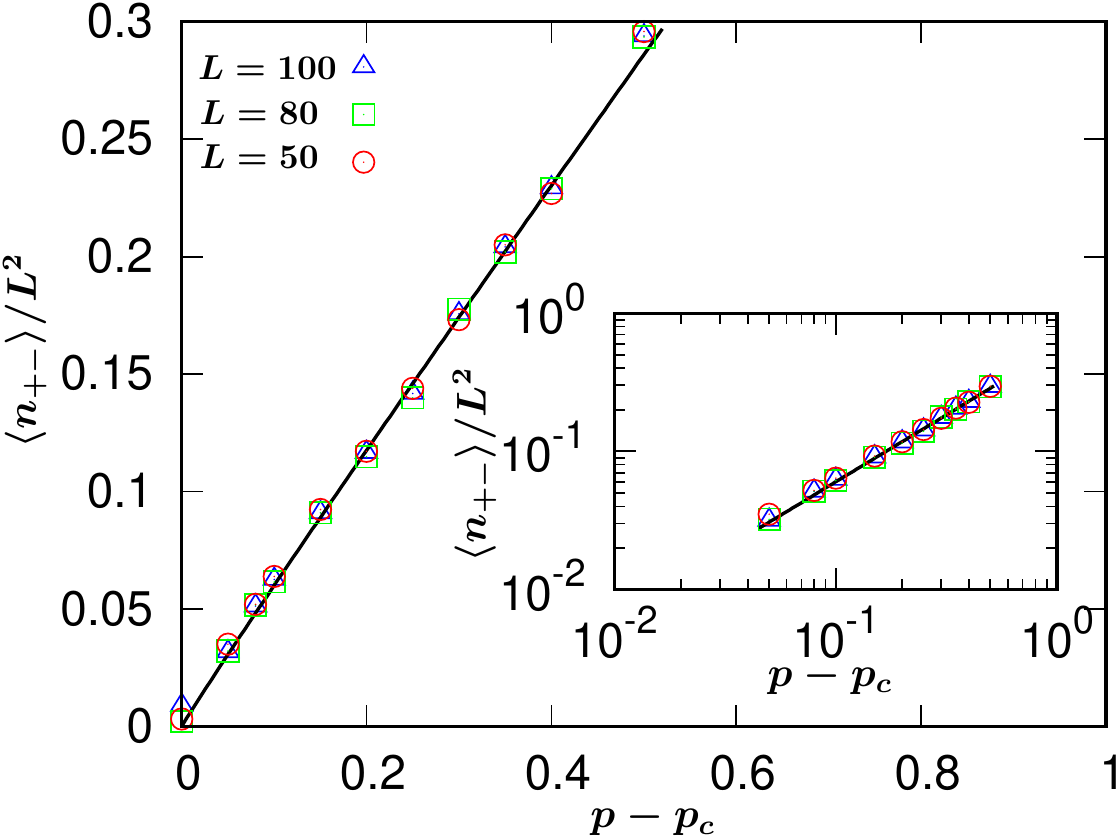}
 \caption{The value of $\langle n_{+-} \rangle$ as a function of
   $p-p_c$ for $L=50,80$ and $100$. The solid line goes as $\sim
   (p-p_c)^{0.97}$. The inset is a log-log plot of the data. It
   suggests that when $p\rightarrow p_c^+$, $\langle n_{+-}
   \rangle\rightarrow 0$, then most of the bonds are activated so that
   nearly all spins are unlikely to flip, resulting in the super slow
   dynamics of the system.}
  \label{fig_bond}
\end{figure*}

In our simulations, we have performed $100$ independent samples to
obtain the number of `unhappy' bonds.  The measured values of $\langle
n_{+-} \rangle$ at the critical temperature can be found in
Fig. \ref{fig_bond}, with a log-log plot as an inset. In particular,
numerically we find that
\begin{equation}
\langle n_{+-}\rangle / L^2 \sim (p-p_c)^{0.97\pm 0.03} \quad \text{for} \quad p\geq p_c.
\end{equation}

When $p\rightarrow p_c^+$, the values of $\langle n_{+-}\rangle$
reduces to zero (or a value close to zero), which means that most of
the active bonds are `happy' so that spins are unlikely to flip.  This
might explain why the system is getting super slow when $p$ approaches
the percolation threshold.

\end{document}